# Further comment on 'Encoding many channels on the same frequency through radio vorticity: first experimental test'

**Michele Tamagnone**[1], **Christophe Craeye**[2] **and Julien Perruisseau-Carrier**[1,3]

[1] Adaptive MicroNanoWave Systems Group, LEMA/Nanolab, Ecole Polytechnique Fédérale de Lausanne (EPFL), 1015 Lausanne, Switzerland
[2] Université Catholique de Louvain (UCL), ICTEAM Institute. Bâtiment Stévin - 2, Place du Levant B-1348 Louvain-la-Neuve, Belgique
[3] E-mail: julien.perruisseau-carrier@epfl.ch

**Abstract**: We show that the reply by Tamburini *et al* (2012 *New J. Phys*. **14** 118002) to our previous comment (2012 *New J. Phys*. **14** 118001) on the experiment reported in (2012 *New J. Phys*. **14** 033001) actually does not invalidate any of the issues raised in our initial comment.

The actual use of OAM modes for increasing –and supposedly without limit– the spectral efficiency of far-field wireless links proposed by Tamburini *et al* [1] has been questioned by various groups [2–4]. The publication of our comment [2] led to a reply [5]. The new claims found in [5] unfortunately require an additional comment to highlight that the objections raised in the reply by Tamburini *et al* [5] do not contradict the conclusions in [2] and do not answer any of the issues raised. In the remainder of this text we briefly expose the shortcomings of [5], without repeating the findings of [2, 3] for brevity.

As a first and important consideration, given the frequency and power range of the Venice experiment described in [1], the underlying physics can be completely described by classical physics, namely here Maxwell's equations together with the constitutive equations of the involved materials. Hence quantum-mechanical arguments [5] that lie beyond classical physics are irrelevant to discuss the concept and results of the experiments reported in [1].

Therefore the general Multiple Input Multiple Output (MIMO) theory, which has been the subject of intense research since the early nineties in the radio communications community, does apply here since it does not make any assumption on the antenna shape or on the particular propagation mode. Indeed, any system with more than one antenna in transmission (Tx) and in reception (Rx) can be regarded as MIMO and described with a channel matrix *H* which, together with the noise at the receiver, determines the transmission capacity of the radio link [3, 6]. In particular the statement "MIMO is a transmission–reception enhancement technique based entirely on the manipulation of linear momentum and polarization" of the reply [5] is not correct as the only assumption of the MIMO general theory on the channel is linearity, which enables the *H* matrix representation; obviously the issue of polarization is very well understood –and exploited– by antenna and wireless engineers, including in MIMO theory, and is not related to the present controversy. In conclusion, the system proposed in [1] can be rigorously considered as MIMO as already explained in [2, 3], and thus cannot overcome well-known MIMO spectral efficiency limits.

As explained in [2], sharp variations of the fields in the receive plane are necessary to transmit multiple channels (we of course exclude the well-known means of using two polarizations). It is possible to prove the finite spatial derivative of the transmitted electric field at the fixed size Rx antenna will always cause an additional attenuation factor for any mode characterized by a null field in the propagation direction (this is the case for the OAM modes other than $l = 0$), which prevents the exploitation of this technique for point-to-point line-of-sight far-field links. This conclusion is in agreement with the high attenuations for OAM

Further comment on 'Encoding (...) through radio vorticity: first experimental test'

transmissions found in [3] and confirmed by private communication with the earlier inventor of the OAM-generating modified dish [7] as used in the experiment of [1].

Tamburini et al [5] denied these findings following an argument that can be summarized as follows:
1) The intensity of angular momentum falls of as $r^{-2}$ in far field just like the EM power density (linear momentum).
2) Hence, taking an integration surface that *grows with the radius*, its area grows as $r^2$ and both the total EM power and the total angular momentum will be constant with r.
3) Citing [5]: "Since angular momentum has the same far-zone behavior [as the linear momentum], angular momentum can also be used for wireless information transfer over huge distances"

Points 1) and 2) are obviously correct, while point 3) is conceptually wrong. Indeed, this argument is valid only under the hypothesis that the integration surface (i.e. the size $d$ of the receiving set of antennas in the Venice experiment) *grows with the distance r*. In our Comment [2] we show that point 3) is not valid when the size $d$ of the OAM receiver *is left constant* while varying the distance $r$, which is of course the practical case of interest. In fact, considering a *fixed size* OAM receiver placed in the far field and along the OAM mode axis, its solid angle as seen by the OAM transmitter will be inversely proportional to the distance squared. This implies that the received power is proportional to $r^{-2}$ for a standard ($l = 0$) uniform phase mode, but the same cannot be stated about OAM modes with $l \neq 0$, since their power density goes to zero in proximity of the OAM phase singularity (i.e. the OAM modes 'axis'). Figure 1 provides an intuitive graphic representation of the origin of the additional attenuation factor that appears in the OAM radio link: in far field, for increasing distances $r$, the OAM Rx antenna "sinks" in the low field area (in blue) surrounding the OAM singularity, causing a total $r^{-4}$ power decay (for $l = 1$).

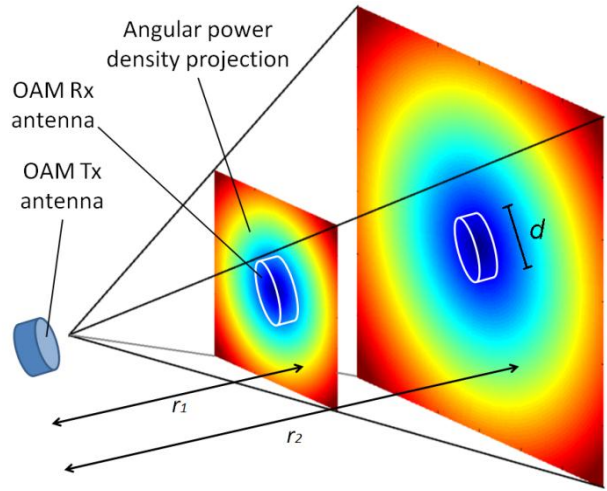

Figure1: Graphical representation of the power collected by an OAM Rx antenna (or OAM detector). For different values of the distance *r*, the 2D angular power density projection is scaled proportionally to *r*, but the size of the Rx antenna is unchanged, and so it "sinks" in the low field area (blue) around the singularity. This effect leads to the additional attenuation factor

As a mathematical support to the above argument, it is worth mentioning that, for Laguerre-Gaussian beams which are carrying OAM, the far electric field for small angles θ from the OAM singularity axis direction depends on r and θ as [8]:

$$|E| \propto \frac{1}{r}\left(\frac{\sqrt{2}\pi\theta W_0}{\lambda}\right)^{|l|} = \frac{1}{r} \cdot \frac{1}{r^{|l|}}\left(\frac{\sqrt{2}\pi d W_0}{\lambda}\right)^{|l|} = \frac{1}{r^{|l|+1}}\left(\frac{\sqrt{2}\pi d W_0}{\lambda}\right)^{|l|}$$

where $\theta = d/r$ for small $\theta$, and where $W_0$ is source aperture and λ the wavelength. For constant $d$, this formula leads to a total attenuation factor of $r^{-(|l|+1)}$ for the electric field, and a total attenuation of $r^{-2(|l|+1)}$ for the power. In particular, the usual $r^{-2}$ free space power attenuation is found for uniform phase modes ($l = 0$) and $r^{-4}$ for $l = \pm 1$ as in the Venice experiment [1], in agreement with the above claims and [2]. Higher order OAM modes lead to increasingly large exponents, requiring even larger distances *d* between the receiving antennas. This also corresponds to the phase diagrams shown in [3], where increments of *l* lead to proportional reduction in the logarithm of the received power in the far field.

In conclusion, the OAM multiplexing scheme presented in [1] can be used for large Tx-Rx distances *r* only if the distance *d* between the Rx antennas is increased proportionally to *r*, as stated explicitly in Section 1 of [5]: "In fact, in all our experiments the field and its phase were measured at varying baseline distances *d* between the two interferometer antennas".

If the distance *d* between the Rx antennas *is kept constant*, the aforementioned additional power decay factor will prevent any capacity gain in the far field. If instead it is *increased proportionally* to the Tx-Rx distance *r*, then the same result can be obtained with the alternative setup proposed in [2], where two standard transmitting antennas are used instead of the modified paraboloid [1,7], showing that OAM modes are not required for such a result.

Finally, as known from basic communication theory, these considerations are completely independent on the modulation [9] used to encode information on the available channel.

**Acknowledgements**: The authors warmly thank Julian Trinder, Ove Edfors, and Laszlo Kish for the fruitful discussion and Prof. George Eleftheriades for pointing toward the link between OAM modes and Laguerre-Gaussian beams.